\documentclass[journal=jacsat,manuscript=article]{achemso}
\usepackage[version=3]{mhchem} 
\usepackage[T1]{fontenc}       
\usepackage{graphicx}
\usepackage{achemso}
\author{S.Erfanifam}
\email{s.erfanifam@gmail.com}
\author{S. M. Mohseni}
\email{m-mohseni@sbu.ac.ir}
\author{L. Jamilpanah}
\author{M. Mohammadbeigi}
\affiliation[Shahid Beheshti university] {Department of physics, Shahid Beheshti University, Tehran, G. C., Evin, Tehran, 19839, Iran}
\author{P. Sangpour}
\affiliation[Institute for materials and energy]
{Department of nanotechnology and advanced materials, Materials and Energy Research Center, Karaj, Iran}
\author{S. A. Hosseini}
\author{A. Iraji Zad}
\affiliation[Sharif university of technology]
{Faculty of physics, Sharif university of Technology, P.O.Box 11155-9161,Tehran, Iran}
\title [An \textsf{achemso} demo]{Tuneable bandgap and spin-orbit coupling by composition control of MoS$_{2}$ and MoO$_{x}$ (X=2 and 3) thin film compounds}
 
\begin{document}

\begin{abstract}
We report on composition controlled MoS$_{2}$ and MoO$_{x}$ (x=2 and 3) compounds electrodeposited on Flourine dopped Tin Oxide (FTO) substrate. It was observed that the relative content has systematic electrical and optical changes for different thicknesses of layers ranging from $\approx$20 to 540 nm. Optical and electrical bandgaps reveals a tuneable behavior by controlling the relative content of compounds as well as a sharp transition from p to n-type of semiconductivity. Moreover, spin-orbit interaction of Mo 3d doublet enhances by reduction of MoO$_{3}$ content in thicker films. Our results convey path-way of applying such compounds in optoelectronics and nanoelectronics devices. 
\end{abstract}
\textbf{keywords}: Low dimensional dichalcogenide compounds, X-ray photoelectron spectroscopy, optical and electrical bandgap

In the past decade, low dimensional binary compounds with general formula MX$_{2}$, has attracted significant attention due to some unique properties of them \cite{Mattheis,Mak,Coehoorn,Yin}. In these compounds M is usually a transition metal and X is a chalcogenide. Among them probably Molybdonium disulfide (MoS$_{2}$) is most well-known material. MX$_{2}$ compounds has potential applications in nanoelectronics, photoelectrochemistry, energy storage and etc\cite{he,huang2015,Klinovaja2013,Bandyopadhyay2015,zhang2015,yufei,yoffe1969,patil}. However, in a form of single MX$_{2}$ compound, bandgap can be tuned by back-gate control or by impurity doping, but it technically requires some additional preparation steps \cite{chih}. Hybrid configuration of MX$_{2}$ compounds can overcome to this drawback. By appropriate combination of these binary compounds a broad range of bandgaps can be obtained which are applicable in optoelectronics and spintronic devices. 

Various techniques such as chemical vapor deposition \cite{zhan} or mechanical (liquid phase) exfoliation have been employed for fabrication of transition metal dichalcogenide compounds, but there is few attempts to fabricate these materials by electrodeposition \cite{ghosh2013,ponomarev,murugesan}. This technique can provide low cost and high quality structures with controllable size made of various compounds. During the fabrication process of MoS$_{2}$ from Mo oxide compounds there are naturally MoO$_{x}$ products; results in MoS$_{2}$/MoO$_{x}$ composition.

In order to exploit the electrical and optical properties of MoS$_{2}$/ MoO$_{x}$ (x=2 and 3) composition at first we refer to the properties of them independently. The bulk MoS$_{2}$ crystal has a trigonal prismatic structure with an inversion symmetry in the middle of the two MoS$_{2}$ monolayers. Unlike, the monolayer and odd number of layers in which the inversion symmetry is absent, the bilayer MoS$_{2}$ and even number of layers has inversion symmetry. In the monolayer structure, there is a relatively (compared to the other transition metals) strong spin-orbit splitting (SOS) in Mo 3d orbitals arisen from lack of inversion symmetry. The SOS in the bilayer MoS$_{2}$ mostly comes from inter-layer coupling. The strong SOS can lead to some important excitonic effects sas well as splitting in valence and conduction band. The SOS value reported in different references is around 150 meV.

MoO$_{2}$ is an unusual metal oxide which exhibits metal-like electrical conductivity (semimetal or a wide band gap n-type semiconductor in bulk)\cite{Moosburger,eyert2000}. During last decade, optical and structural properties of this compound has been thoroughly investigated. MoO$_{2}$ is a mixed ionic electronic conductor \cite{harrison2015} and its electrical resistivity is very controversial. Some reports indicate conductivity of MoO$_{2}$ at room temperature  in a broad range from $\mu$ to m  $\Omega$.cm depending on crystallization\cite{smith}. Some news shows a slightly correlated metallic conduction \cite{harrison2015}. SOS value of this material is less than that of MoS$_{2}$ at which is around 130 meV. The bulk MoO$_{2}$ crystal has a monoclinic structure. Morphology and preparation method can affect its optical and electrical properties. This compound has mixed valence band similar to that in MoS$_{2}$ in which the orbitals in valence band are mixed and therefore can transport both of electron and hole. Technically this characteristic can be important in various applications. 

Molybdenum trioxide, with an orthorhombic symmetry is intrinsically an n-type semiconductor with 3.2 eV band gap. It has been widely investigated due to its high work function \cite{guo}. MoO$_{3}$ is a transparent material that is capable to Li$^{+}$ and H$^{+}$ be intercalated \cite{Balendhran}. These properties make it a promising candidate for photorechargable and photoelectrochemical applications\cite{lou,lou2014,yao2012}.

The present paper explores the structural and optical properties of electrodeposited MoS$_{2}$, MoO$_{2}$ and MoO$_{3}$ compounds made on transparent FTO substrate. However, according to Ref.\cite{ghosh2013} a transition metal substrate is required to act as catalyst to have MoS$_{2}$, but in this research we show that it is possible to have a composition of these compounds on a transparent substrate of FTO. 
We provide a fabrication recipe to control each of these MoS$_{2}$/MoO$_{2}$/MoO$_{3}$ compounds and report on their interesting characteristics. A systematic growth of these compounds show transition from p to n semiconductivity with increasing thickness.
\begin{figure}
\begin{center}
\includegraphics[scale=0.22]{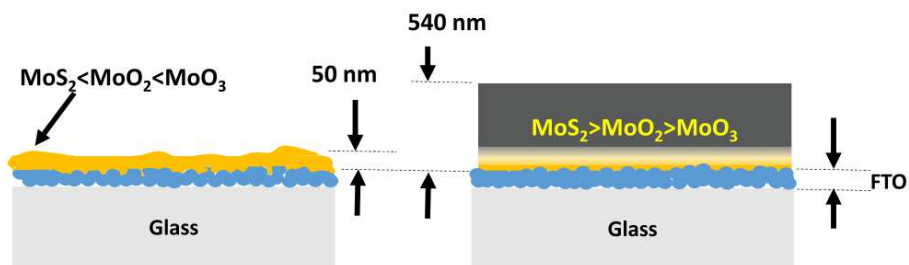}
\end{center}
\vspace{-15pt}
\caption{(Color online) Schematic representation of two distinct samples with 50 and 540 nm thickness. Morphology and relative content change is shown qualitatively by color gradient at thinner and thicker samples relatively.}
\label{sampleg}
\end{figure}
%
\section{Results and discussion}

We study electrodeposited MoS$_{2}$/MoO$_{2}$/MoO$_{3}$ compounds in different thicknesses by changing the deposition time on the FTO substrate. Our observations shows that the thick layers are black while thinner ones, below 1 min deposition time, are amber yellow and more transparent.
\begin{figure}
\begin{center}
\includegraphics[scale=0.35]{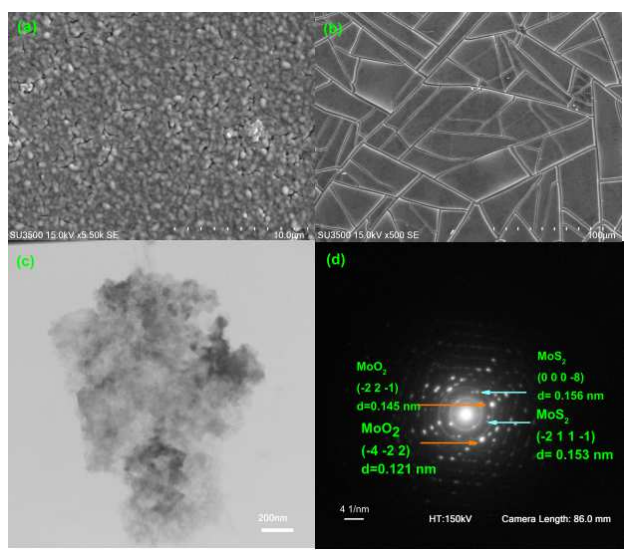}
\end{center}
\vspace{-15pt}
\caption{(Color online) SEM micrograph of thin (a) and thick (b) layers exhibiting morphology as well as substrate effect. TEM (c) and SAED patterns (d) of the thicker layer. The arrows point to the compound, crystallographic direction and distance of the planes.}
\label{saed2}
\end{figure}
As it can be seen from Fig.\ref{sampleg}, morphology and electrical as well as optical properties of electrodeposited thin layers that will be shown later are strongly depend on the composition. However, we expect the interplay between H$^{+}$ and Oxygen vacancies can affect on the composition and charge carrier doping on the layer.

Fig.\ref{saed2}a and b show SEM micrograph of the as-deposited samples with different deposition times of 30 s and 10 min respectively. It is clearly seen that at thinner layers the roughness of substrate affects the morphology of the layer, While thick layer shows large-scale flat islands. Fig. \ref{saed2}c and d exhibit TEM and selected area diffraction pattern of scratched thick layer. Interestingly, we observed that the fabricated material is mostly polycrystalline. The analysis by Diffraction Process software detects different phases which are attributed to the MoO$_{2}$ and MoS$_{2}$.


\begin{figure}
\begin{center}
\includegraphics[scale=0.45]{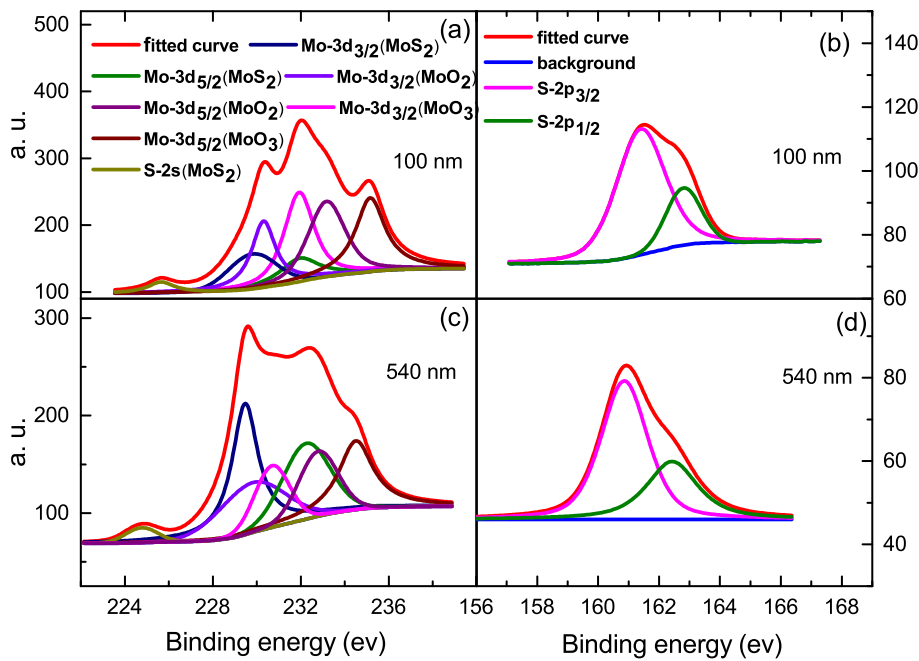}
\end{center}
\vspace{-15pt}
\caption{(Color online) X-ray photoelectron spectra of MoS$_{2}$, MoO$_{2}$ and MoO$_{3}$ compounds at 100 nm and 540 nm thickness of layer.}
\label{xps4}
\end{figure}

In order to understand the chemical composition and accuratly analysis of the observations, XPS (X-ray photoelectron spectroscopy) spectra of 30s and 10 min as-deposited samples (shown in Fig. \ref{xps4}) is acquired. The obtained XPS data from 0 to 1200 eV and convoluted to analyze the XPS peaks. Best fitted Gaussian curves were used to determine the peak positions presented in table \ref{relative}. noted that, C 1s calibration peak (284.8 eV) was employed to find the correct binding energy of all peaks. For 100 nm sample, we have a shift in peaks towards higher energies because of formation of more MoO$_{2}$ and MoO$_{3}$. 

\begin{table}[ht]
\caption{ Relative content of compounds in as-deposited samples.}
\vspace{.25cm}
\centering
\begin{tabular}{c c c c}
\hline\hline
\textbf{Compound} & \textbf{MoS$_{2}$} & \textbf{MoO$_{2}$} & \textbf{MoO$_{3}$}\\ [0.8ex]
\hline 
Relative content (100 nm) & 0.20 & 0.36 & 0.43\\ 
\hline
Relative content (540 nm) & 0.43 &  0.29 & 0.28\\
 [2ex]
\hline
\end{tabular}
\label{relative}
\end{table}

From table.\ref{relative}, it is seen that the relative content of MoS$_{2}$ has noticeable increase from 20 percent to 43 percent. This enhancement comes from less Oxygen evolution as well as disappeared-substrate effect for thicker layers. However, the chemical stability of MoO$_{3}$ is highest among these three compound, the content of that decreases at higher deposition times and thicknesses. The relative content evolution versus layer thickness is schematically shown in Fig.\ref{sampleg}. 

\begin{table}[ht]
\caption{Binding energy of Mo and S in as-deposited samples. Summary of the peak positions, spin-orbit coupling and related shifts in the Mo 3d and S 2p for 100 nm and 540 nm films.}
\vspace{.25cm}
\centering
\begin{tabular}{c c c c c c c}
\hline\hline
\textbf{100 nm} & Mo 3d$_{3/2}$ & Mo 3d$_{5/2}$ & $\Delta_{SOC}$(Mo 3d) & S-2p$_{1/2}$ & S-2p$_{3/2}$ & $\Delta_{SOC}$(S 2p)\\ [0.05ex]
\hline 
MoS$_{2}$ & 231.97 & 229.98 & 2.15 & 161.40 & 162.80 & 1.38\\ 
\hline
MoO$_{2}$ & 233.18 & 230.30 & 2.85 \\
\hline
MoO$_{3}$ & 235.15 & 231.96 & 3.28 \\
\hline
\textbf{540 nm} &  &  \\ [0.05ex]
\hline 
MoS$_{2}$ & 232.24 & 229.47 & 2.87 &160.86 & 162.43 & 1.53\\ 
\hline
MoO$_{2}$ & 232.79& 229.81 & 2.77\\
\hline
MoO$_{3}$ & 234.50 & 230.69 & 3.74 \\
[2ex]
\hline
\end{tabular}
\label{binding}
\end{table}

The peak positions of Mo and S for two different thicknesses are shown in table \ref{binding}. The peak position of oxygen is not shown due to the substrate effect and not possible accurate conclusions. Decreasing of the binding energy by increasing MoS$_{2}$ content in thicker layers shows higher charge density in spite of reduction of oxidation states.

\begin{figure}
\begin{center}
\includegraphics[scale=0.45]{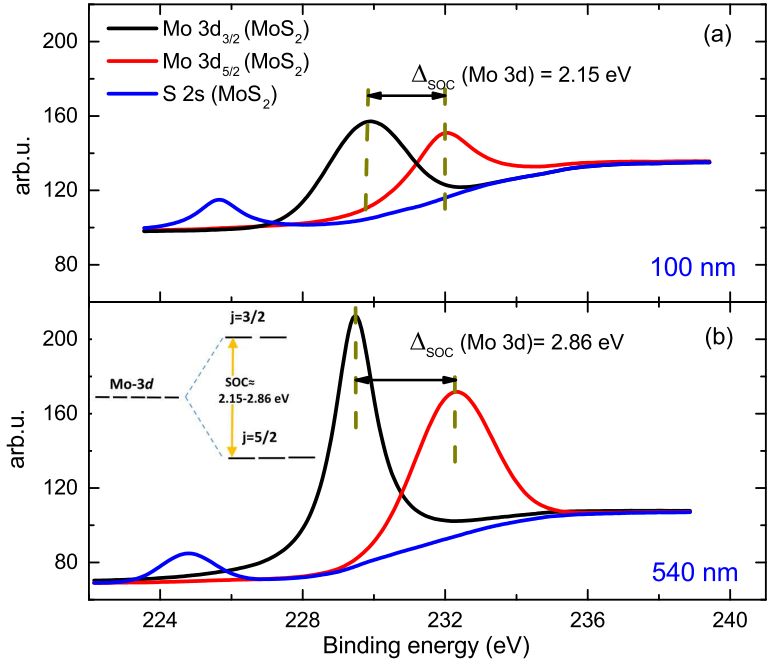}
\end{center}
\vspace{-15pt}
\caption{(Color online) Spin-orbit coupling of Mo 3d electrons in MoS$_{2}$ for two different thickness of layer. Moreover, 2s peak of sulfur exhibits a significant redshift towards low binding energy. Inset in b shows schematic representation of the spin orbit coupling of L=2 and S=1/2 in 3d orbitals.}
\label{soc}
\end{figure}
Fig.\ref{soc} shows spin-orbit coupling (SOC) of Mo 3d$_{3/2}$ and Mo 3d$_{5/2}$ for MoS$_{2}$ in 100 nm and 540 nm samples which exhibits increasing from 2.15 eV to 2.86 eV. At first approximation, this effect can be explained by relative increasing of MoS$_{2}$ content. The observed SOC in our experimental data is in agreement with reported results \cite{Senthilkumar}. 
In the monolayer or bulk MoX$_{2}$ (x=O and S) compounds, depending on the odd or even number of layers (SOS), arisen from inversion symmetry breaking or interlayer coupling, can lead to the valence and conduction band splittings at first Brllouin zone. The Key point is that this splitting is mostly depending on SOC coming from the presence or absence of inversion symmetry and inter-layer coupling rather than SOC. The correlation between SOC and SOS is not very clear yet. The excitonic peaks arisen from this splitting is usually shown in the UV-Vis experiments as we will discuss in the next section. Apparently, in our composition the valence bande splitting is not observed in UV-Vis results. With this, one can conclude that the presence of MoO$_{2}$ and MoO$_{3}$ suppress valence band splitting. 

According to the calculated SOC of Mo 3d orbitals in MoO$_{x}$ (x=2 and 3) and  MoS$_{2}$ presented in table. \ref{binding}, this parameter value in MoO$_{3}$ steps up from 3.28 to 3.74  eV, while in case of MoO$_{2}$ it steps down from 2.85 to 2.77 eV. This values are reasonable according to this fact that the SOC has direct relation with molecular mass and molecular mass of MoO$_{3}$ is higher than MoO$_{2}$. In addition, we observed that 2p peaks of sulfur in MoS$_{2}$ increases from 1.4 to 1.55 eV. These values are in accordance with previous reports with some small deviations. However, in spite of the MoS$_{2}$ the obtained values for MoO$_{3}$ is slightly higher than the reported values.
\begin{figure}
\begin{center}
\includegraphics[scale=0.5]{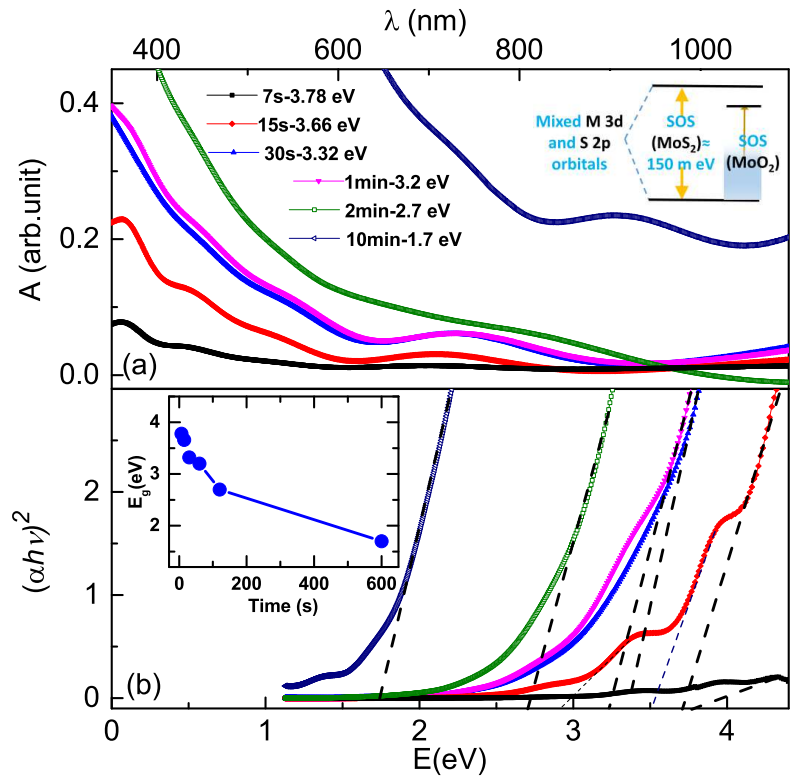}
\end{center}
\vspace{-15pt}
\caption{(Color online) a: UV-visible spectrum of as deposited thin films with different deposition times. b: Tauc plot and bandgap (inset). The broad absorption band centered at 416 nm (2.98 eV) arising from the complicated ineraband transitions is also observed. However the excitonic peaks of spin orbit and interlayer couplings are not well resolved. The inset shows screening effect of MoO$_{2}$ valence band splitting on MoS$_{2}$ schematically.}
\label{uv}
\end{figure}
%
In the optoelectronics and material science one of the powerful and common characterization techniques of transparent materials is frequency dependence absorption measurement by UV-Vis spectroscopy. Fig.\ref{uv}a shows the UV-Vis spectrum for different deposition times and thicknesses. By probing from 300 to 1100 nm, interestingly we observed various anomalies and peaks. These anomalies can be attributed to interaband and interband transitions. All peaks have redshift for thinner layers. The region that we expect to observe excitonic peaks is not exactly coincide to the reported values of 623 and 683 nm for MoS$_{2}$and neither very well separated from each other. However, high energy peaks related to interband transition are in agreement with reported results. 
According to the XPS results, the explanation of absent double excitonic peaks in the visible range of data is in screening of the valence band splitting in MoS$_{2}$ by SOC in MoO$_{2}$. Another scenario is that, the energy separation of Mo 3d degeneracy-lifted orbitals (due to the SOS coming from inversion symmetry breaking or inter-layer coupling) in both compounds are beyond the resolution of our UV-Vis instrument (see inset of Fig. \ref{uv}.a) and it is shown as a broad peak at the expected frequency region. However, in our composition the MoO$_{3}$ is also present, but we expect no contribution due to no excitonic effects in this material.
\begin{figure}
\begin{center}
\includegraphics[scale=0.5]{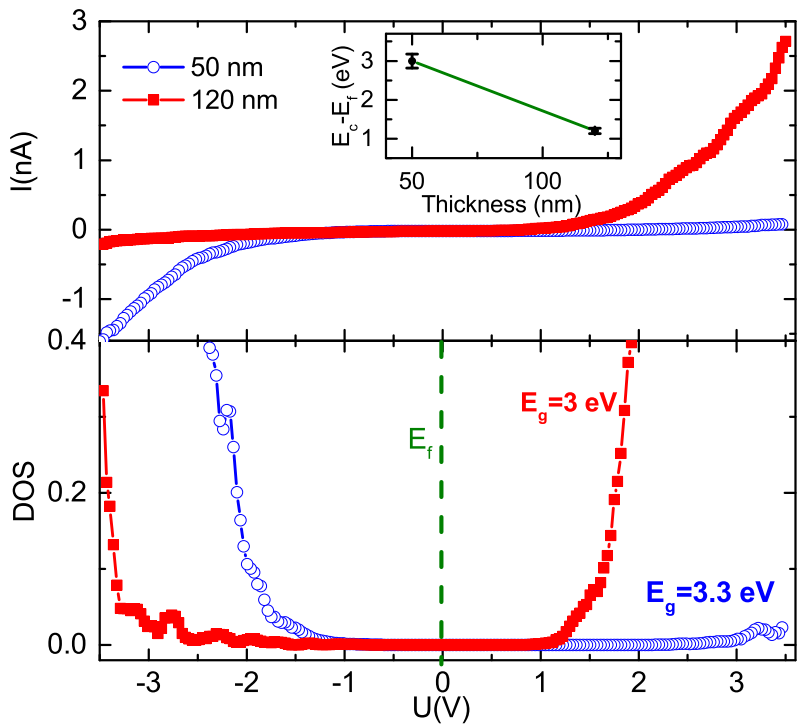}
\end{center}
\vspace{-15pt}
\caption{Scanning tunneling spectroscopy and DOS of thick and thin films. Inset shows E$_{c}$-E$_{f}$ change versus layer thickness.}
\label{sts}
\end{figure}
Our UV-Visible results are in agreement with reference \cite{bojana2011}. In addition, our results agrees well in sense of trend from visible to ultrviolet with reference\cite{sreedhara}. We expect the band structure change to be arisen from MoO$_{2}$ and MoO$_{3}$ effect in texture of MoS$_{2}$. It is worth to note that by comparison of insets in Figs.\ref{uv} and \ref{soc}, it can be concluded that the SOS and SOC have two different origin.

Usually, the optical bandgap of semiconductor materials can be calculated by plotting the ($\alpha$h$\nu$)$^{2}$ versus h$\nu$ in Uv-Vis spectrum where $\alpha$ is absorption coefficient (Tauc plot). Fig.\ref{uv}b shows  Tauc plot for extraction of indirect bandgap obtained from UV-Visible spectrum of the thin films for different thicknesses. Our calculations indicates 3.78 eV energy gap for thinnest layer with 7 S deposition time, while the bandgap for 10 min deposited one is 1.77 eV. Complementary experiments for the times, between these two values, exhibit a systematic and tuneable bandgap change. From this, one can conclude that this change is not affected strongly from dislocations or impurities. 

Interestingly, we observed an unusual step-like behavior below 15 s deposition time. However one can speculate that the different composition due to substrate effect at very thin layers can play important role. We believe that such behavior can be attributed to the curved shape of layer which is formed on top of rough FTO nanoparticles. This bending on layer cause lattice distortion and affects on the band structure. Our extended experiment at 5V and 7.5V (not shown here)showed no significant change on these features except the steps are going to be smeared out.
Comparison of optical bandgap and electrical bandgap can give us detailed understanding from carrier correlations and charge density in various compositions. Fig. \ref{sts} exhibits scaninng tunneling spectroscopy results of two samples with 50 and 120 nm. Interestingly, we observed that by increasing the thickness band gap decreases from 3.3 eV in thinner layer to 2.8 eV in thicker layer with 540 nm thickness. According to our measurements the optical and electrical bandgaps are comparable and the difference is not significant. In addition, Fermi energy gets closer ( from 2 to 1.2 eV) to conduction band which indicates changing of semiconductor type from p to n. The origin of this behavior is in composition change as discussed in XPS results. However, proton intercalation in thinner layers can play important role and our explanation is following. 

At the beginning of deposition, due to the catalytic effect of the substrate the protons penetrates and forms a bound state with  Oxygen  atoms  to  form  water  vapor  that leaves  Oxygen  vacancies  behind. This vacancies make the semiconductor p-type. After a while and increasing the distance of solution layer interface from FTO substrate, the substrate effect smears out and Oxygen evolution decrease and charge density arisen from n-type semiconductor dominated. 

\section{Conclusions}

In conclusion, MoS$_{2}$, MoO$_{x}$(x=2 and 3) composition in the form of thin films are electrodeposited on a transparent (FTO) substrate with different thicknesses. Optical and electrical characteristics show a tunable composition. This tunability enables us to control the optical bandgap and the spin-orbit doublet splitting SOS which can be useful for various applications ranging from  photo-electrocatalyst and supercapacitors to spintronic devices. The microscopical explanation of this behavior is still enigmatic and requires more theoretical and experimental explorations. We propose that the proximity effect of MoO$_{x}$(x=2, 3) on MoS$_{2}$ in our composition can be responsible of suppression of valence band splitting in MoS$_{2}$ or  positioning of SOS energy level of MoO$_{2}$ in between SOS energy levels of MoS$_{2}$ makes the excitonic peaks undetectable. Moreover, at very thin layers below 50 nm mid gap states emerges that comes from bending of layer over FTO nanoparticles and changing lattice parameters by distortion. 
we believe that this research is a prominent example of tuning of magnetic and electronic properties of hybrid dichacogenide compounds based on composition control.
\section{Methods}

DC electrodeposition was carried out to make thin layers from an electrolyte containing sodium molybdate (Na$_{2}$Mo$_{2}$O$_{4}$) (0.5 M) with disodium sulphide (Na$_{2}$S.5H$_{2}$O) (30gr/l) similar to Ref. \cite{ghosh2013}. The pH value of the electrolyte was adjusted to 7 by adding dilute sulphuric acid. A two electrode potentiostat having Pt as anode and FTO substrate as cathode was employed for the film growth. 
In a series of samples the deposition was carried on different deposition times between 7 s and 10 min. Applied DC voltage was adjusted to 2.5 V in which the current density during the deposition kept at around 2 mA/cm$^{2}$, well below 10 mA/cm$^{2}$ in accordance with previous reports to avoid highly reduction of MoO$^{2+}$ ions. Electrodeposition was carried out at room temperature.

\begin{acknowledgement}

Salim Erfanifam acknowledges the support from the Iranian Elites foundation. S. M. Mohseni acknowledges ISEF foundation. 

\end{acknowledgement}

\end{document}